\documentclass[12pt]{article}
\usepackage[dvips]{graphicx}
\usepackage{amsmath}
\usepackage{bm}

\begin{document}
\begin{flushright}
  OU-HET-644 \ \\
\end{flushright}
\vspace{0mm}

\begin{center}
\large{Chiral Quark Soliton Model and Nucleon Spin Structure Functions}
\end{center}
\vspace{0mm}
\begin{center}
M.~Wakamatsu\footnote{Email \ : \ wakamatu@phys.sci.osaka-u.ac.jp}
\end{center}
\vspace{-4mm}
\begin{center}
Department of Physics, Faculty of Science,
Osaka University, \\
Toyonaka, Osaka 560-0043, JAPAN
\end{center}

\centerline{\bf Abstract}

\vspace{2mm}
\begin{small}
The chiral quark soliton model (CQSM) is one of the most successful models
of baryons at quark level, which maximally incorporates the most important
feature of low energy QCD, i.e. the chiral symmetry and its spontaneous
breakdown. Basically, it is a relativistic mean-field theory with
full account of infinitely many Dirac-sea quarks in a
rotational-symmetry-breaking mean field of hedgehog shape.
The numerical technique established
so far enables us to make a nonperturbative evaluation of Casimir
effects (i.e. effects of vacuum-polarized Dirac sea) on a variety of
baryon observables. This incompatible feature of the model manifests
most clearly in its predictions for parton distribution functions
of the nucleon. In this talk, after briefly reviewing several basic
features of the CQSM, we plan to demonstrate in various ways that this
unique model of baryons provides us with an ideal tool for disentangling
nonperturbative aspect of the internal partonic structure of the
nucleon, especially the underlying spin structure function of the nucleon.
\end{small}

\vspace{6mm}
\noindent
\begin{large}
{\bf 1. Introduction}
\end{large}

\vspace{3mm}
What is the CQSM like? To answer this question, it is instructive
to ask another simpler question.
What is, or what was, the Skyrme model ? In a word, the
famous Skyrme model is Bohr's model in baryon physics.
The simplest microscpic basis of Bohr's collective model
of rotational nuclei is provided by the deformed Hartree-Fock theory
supplemented with the subsequent cranking quantization.
Very roughly speaking, the relation between the CQSM and the Skyrme
model is resembling the relation between these two theories in
nuclear physics. Let us start with a brief history of the CQSM.

\begin{itemize}
\item The model was first proposed by Diakonov, Petrov and
Pobylitsa based on the instanton picture of the QCD vacuum
in 1988 \cite{DPP88}.

\item In 1991 \cite{WY91}, we have established a basis of numerical
calculation, which enables us to make nonperturbative
estimate of nucleon observables with full inclusion of the deformed
Dirac-sea quarks, by extending the method of
Kahana, Ripka and Soni \cite{KR84},\cite{KRS84}.
Also derived and discussed in this paper is the nucleon spin sum rule,
which reveals the important role of quark orbital angular
momentum in the nucleon spin problem.

\item In 1993, we noticed the existence of novel $1 / N_c$
correction to some isovector observables, which is totally
missing within the framework of the
Skyrme model, but it certainly exists within the CQSM, so that it
resolves the long-standing $g_A$-problem inherent in the hedgehog
soliton model \cite{WW93} (see also \cite{CBGPPWW94}).

\item The next important step is an application of the model to
the physics of parton distribution functions of the nucleon,
initiated by Diakonov et al. \cite{DPPPW96},\cite{DPPPW97} and
also by T\"{u}bingen group \cite{WGR96},\cite{GRW98}.

\end{itemize}

\vspace{5mm}
\noindent
\begin{large}
{\bf 2. Main achivements of the CQSM for low energy observables}
\end{large}

\vspace{3mm}
Skipping the detailed explanation of the model, I just summarize below
several noteworthy achievements of the CQSM for low energy baryon
observables.

\begin{itemize}

\item First of all, it reproduces unexpectedly small quark spin
fraction of the nucleon \cite{WY91},\cite{WK99}
\nocite{WW00}-\cite{Waka03} in conformity
with the famous EMS observation \cite{EMC88} : 
\begin{equation}
 \Delta \Sigma \ \simeq \ 0.35.
\end{equation}
\item Secondly, it reproduces fairly large pion-nucleon sigma-term favored
in the recent phenomenological determination
\cite{OW04} (see also \cite{DPP89}) : 
\begin{equation}
 \Sigma_{\pi N} \ \simeq \ 60 \,\mbox{MeV}.
\end{equation}
\item Furthermore, it resolves the famous $g_A$-problem of the Skyrme model
as \cite{WW93},\cite{CBGPPWW94}
\begin{eqnarray}
 g_A^{(Skyrme)} &=& g_A (\Omega^0) \ + \ g_A (\Omega^1) \ \simeq \ 
 0.8 \ + \ 0.0 \ = \ 0.8 , \\
 g_A^{(CQSM)} &=& g_A (\Omega^0) \ + \ g_A (\Omega^1) \ \simeq \ 
 0.8 \ + \ 0.4 \ = \ 1.2 .
\end{eqnarray}

\end{itemize}

Unfortunately, most baryon observables are quite insensitive
to the differences of low energy models, which results in
masking the potential ability of the CQSM as compared with the others.
It turns out, however, that that the superiority of the CQSM as a
field theoretical model of baryons manifests most drastically
in its predictions for the internal partonic structure of the nucleon.

\vspace{5mm}
\noindent
\begin{large}
{\bf 3. On the role and achievements of CQSM in DIS physics}
\end{large}

\vspace{3mm}
The standard approach to the DIS (deep-inelastic-scattering) physics
is based on the so-called {\it factorization theorem},
which states that the DIS amplitude is
factorized into two part, i.e. the {\it hard part} which can be handled
by the perturbative QCD and the {\it soft part} which contains
information on the nonperturbative quark-gluon structure of the nucleon.
The soft part is usually treated as a blackbox, which should be determined
via experiments. This is a reasonable strategy, since we have no
simple device to solve nonperturbative QCD.
We however believe that, even if this part is completely fixed by experiments, one still wants to know why those parton distribution
functions (PDFs) take the form so determined !
Nonstandard but complementary approach to DIS physics is necessary here to
understand hidden chiral dynamics of soft part, based on models or
on lattice QCD.

There are several merits of the CQSM over many other effective model of baryons. First, it is a relativistic mean-field theory of quarks, consistent with
the large $N_c$ QCD supplemented with the $1 / N_c$ expansion.
Secondly, the field theoretical nature of the model, i.e. nonperturbative
inclusion of polarized Dirac-sea quarks, enables reasonable estimation
not only of quark distributions but also of {\it antiquark} distributions.
Finally, only 1 parameter of the model, i.e. the dynamical quark mass $M$,
was already fixed from low energy phenomenology, which means that we can make {\it parameter-free} predictions for parton distribution
functions. As a matter of course, the biggest default of the model
is the lack of the explicit gluon degrees of freedom.

\begin{figure}[!h] \centering
\begin{center}
 \includegraphics[width=12.5cm,height=16.5cm]{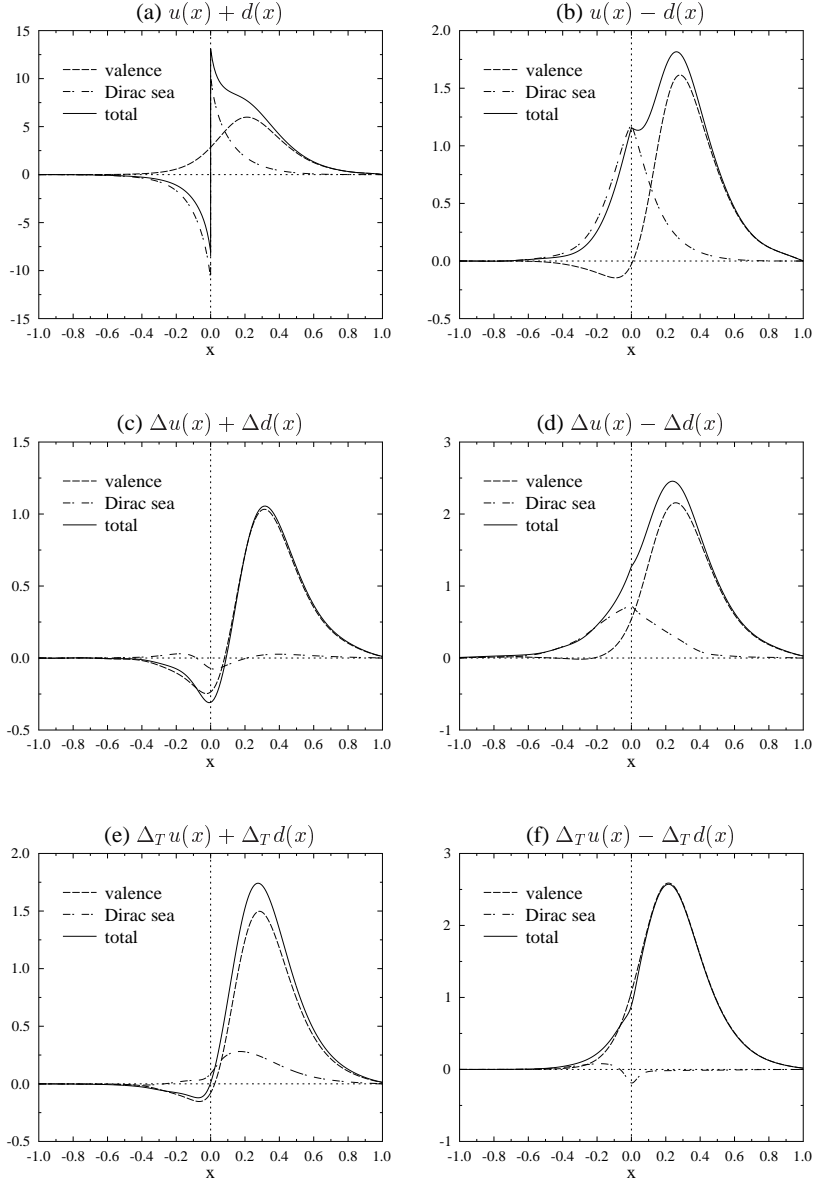}
\end{center}
\vspace*{-0.5cm}
\caption{The CQSM predictions for the fundamental twist-2 PDFs of the
nucleon : isoscalar and isovector unpolarized PDFs ((a) and (b)),
isoscalar and isovector longitudinally polarized PDFs ((c) and (d)),
and isoscalar and isovector transversity distributions ((e) and (f)).}
\label{Fig:Twist2Pdf}
\end{figure}%

In Fig.\ref{Fig:Twist2Pdf}, we summarize parameter-free predictions
of the CQSM for the three fundamental twist-2 PDFs.
They are the unpolarized PDF with
isoscalar and isovector combinations, the longitudinally polarized
PDF with isoscalar and isovector combinations, and finally the
transversities with isoscalar and isovector combinations.
Noteworthy here is totally {\it different behavior} of the Dirac-sea
contributions in {\it different PDFs}.

The crucial importance of the Dirac-sea contribution
can most clearly be seen in the isoscalar unpolarized PDF.
First, I recall that the distribution function in the negative
$x$ region should be identified with the antiquark distribution
with the extra minus sign.
\begin{equation}
 \bar{q} (x) \ = \ - \,q(-x), \hspace{10mm} (0 < x < 1).
\end{equation}

Then, one can see that the positivity of the antiquark distribution
$\bar{u}(x) + \bar{d}(x)$ is satisfied only after including the
Dirac-sea contribution.
It is also seen to generate sea-like soft component in the quark
distribution in the small $x$ region, as required in the GRV analysis
even at the low energy scale \cite{GRV95}.

Turning to the isovector unpolarized PDF, I point out that the
$u(x) - d(x)$ is positive with sizable magnitude in the negative
$x$ region due to the effect of Dirac-sea contribution.
Because of the charge
conjugation property of this distribution, it means that
$\bar{u}(x) - \bar{d}(x)$ is negative or $\bar{d}(x) - \bar{u}(x)$
is positive in consistency with the famous NMC
observation \cite{Waka92}\nocite{WK98}-\cite{PPGWW99}.
One can also confirm that the model prediction for the
$\bar{d}(x)/\bar{u}(x)$ ratio is consistent with the Fermi-Lab
Drell-Yan data at least qualitatively \cite{Waka03}.

Although we do not have enough space to go into the detail,
we can also show that the
model also reproduces all the characteristic features of the
longitudinally polarized structure functions
of the proton, neutron and the deuteron without introducing any
additional parameters \cite{WK99},\cite{Waka03}.

\vspace{10mm}
\noindent
\begin{large}
{\bf 4. Chiral-odd twist-3 distribution function $e(x)$}
\end{large}

\vspace{3mm}
The distribution function $e(x)$ is one of the three twist-3
distribution functions of the nucleon. 
Why is it interesting ?
Firstly, its first moment is proportional
to the famous $\pi N$ sigma term. Secondly, within the framework
of perturbative QCD, it was noticed that this
distribution function may have a delta-function type singularity
at $x=0$ \cite{BK02}.
However, the physical origin of this delta-function type singularity
was left unclear within the perturbative consideration.

\begin{figure}[!h] \centering
\begin{center}
\includegraphics[width=3.0in]{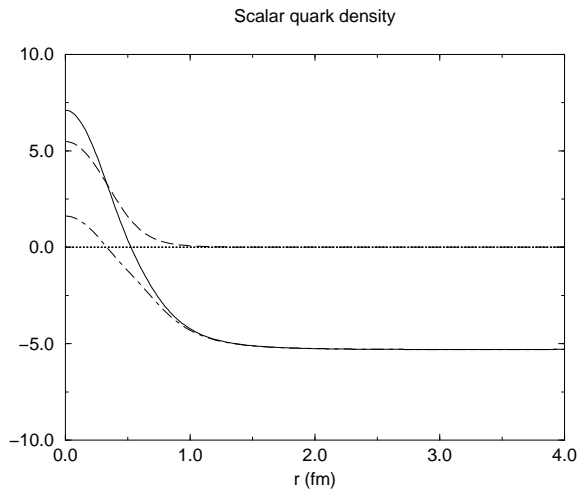}
\end{center}
\vspace*{-0.5cm}
\caption{The scalar quark density predicted by the CQSM.}
\label{Fig:Sdens}
\end{figure}

By utilizing the advantage of the CQSM, in which the effects of
Dirac-sea quarks can be treated nonperturbatively, we have tried
to clarify the physical origin of this delta-function type
singularity \cite{WO03},\cite{OW04}.
We first verified that, because of the spontaneous chiral symmetry
breaking of the QCD vacuum, the scalar quark density of the nucleon
does not damp as the distance from the nucleon center becomes large,
but it approaches a nonzero negative constant, which is nothing but the
vacuum quark condensate. (See. Fig.\ref{Fig:Sdens}.)

It was shown further that this extraordinary nature of the scalar quark
density in the nucleon, i.e. the existence of the {\it infinite range}
quark-quark correlation of scalar type, is the physical origin of
the delta-function singularity in the chiral-odd twist-3 distribution $e(x)$.
This singularity of $e(x)$ will be observed as the violation
of $\pi N$ sigma-term sum rule. To confirm this interesting possibility,
we need very precise experimental information for $e(x)$ through the
semi-inclusive DIS scatterings.

\vspace{10mm}
\noindent
\begin{large}
{\bf 5. Proton spin problem revisited : current status and resolution}
\end{large}

\vspace{3mm}
Now, we come back to our biggest concern of study, i.e. the nucleon spin
problem. Recent two remarkable progresses may be worthy of mention.
First, the quark polarization $\Delta \Sigma$ has been fairly
precisely determined, through the high-statistics measurements of
deuteron spin structure function by the COMPASS and HERMES
groups \cite{COMPASS07},\cite{HERMES07}.
Second, a lot of evidences have been accumulated, which indicate
that the gluon polarization is likely to be small or at least
it cannot be large enough to resolve the puzzle of the missing nucleon
spin based on the $U_A (1)$ anomaly scenario.
A general consensus now is therefore as follows.
About $1/3$ of the nucleon spin is carried by the intrinsic quark spin,
while the remaining $2/3$ should be carried by $L^Q, \Delta g$, and $L^g$.

Recently, Thomas advocates a viewpoint that the modern spin discrepancy
can well be explained in terms of standard features of the nonperturbative
structure of the nucleon, i.e. relativistic motion of valence quarks,
the pion cloud required by chiral symmetry, and an exchange current
contribution associated with the one-gluon-exchange hyperfine
interaction \cite{Thomas08}\nocite{MT08}-\cite{Thomas09}.
His analysis starts from an estimate of the orbital angular momenta
of up and down quarks based on the improved (or fine-tuned) cloudy bag
model taking account of the above-mentioned effects.
Another important factor of his analysis is the observation that
the angular momentum is not a renormalization group invariant quantity,
so that the above predictions of the model should be associated with a
very low energy scale, say, $0.4 \,\mbox{GeV}$.
Then, after solving the QCD evolution equations for the up and down
quark angular momenta, first derived by Ji, Tang and
Hoodbhoy \cite{JTH96}, he was led
to a remarkable conclusion that the orbital angular
momenta of up and down quarks cross over around the scale of
$1 \,\mbox{GeV}$. This crossover of $L^u$ and $L^d$ seems absolutely
necessary for his scenario to hold.
Otherwise, the prediction $L^u - L^d > 0$ of the
improved cloudy bag model given at the low energy scale is incompatible
with the current empirical information or lattice QCD simulations at
the high energy scale, which gives $L^u < 0, L^d > 0$.

\begin{figure}[!h] \centering
\begin{center}
 \includegraphics[width=10.0cm,height=8.0cm]{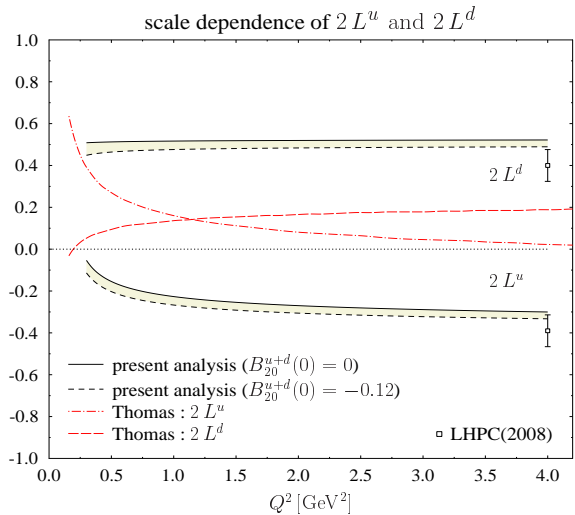}
\end{center}
\vspace*{-0.5cm}
\caption{Our semi-phenomenological predictions of the
orbital angular momenta of up and down quarks in the proton
are compared with the corresponding results of Thomas'
analysis \cite{Thomas09}.
Also shown for comparison are the predictions of the LHPC
lattice simulations for $2 \,L^u$, and $2 \,L^d$ given at
the scale $Q^2 = 4 \,\mbox{GeV}^2$ \cite{LHPC08}.}
\label{Fig:LuLd}
\end{figure}%

On the other hand, we have recently carried out a semi-empirical
analysis of the nucleon spin contents based on Ji's angular momentum
sum rule, and extracted the orbital angular momentum of up and down
quarks as functions of the scale \cite{WN08}. (See also \cite{WN06}.)
Remarkably, we find no crossover of $L^u$ and $L^d$ when $Q^2$ is varied,
in sharp contrast to Thomas' analysis. This difference is remarkable,
since if there is no crossover of $L^u$ and $L^d$, Thomas' scenario
for resolving the proton spin puzzle is seriously challenged.

We show in Fig.\ref{Fig:LuLd} the results of our semi-empirical
analysis for $L^u$ and $L^d$ in comparison with the corresponding
predictions by Thomas.
As already mentioned, Thomas' results show that the orbital
angular momenta of up and down quarks cross over around the scale
of $1 \,\mbox{GeV}$. In contrast, no crossover of $L^u$ and $L^d$
is observed in our analysis : $L^d$ remains to be larger than $L^u$
down to the scale where the gluon momentum fraction vanishes.
Comparing the two, the cause of this difference seems obvious.
Thomas claims that his results are qualitatively consistent with the
empirical information and the lattice QCD data at high energy scale.
(We recall that the sign of $L^{u-d}$ at the high energy scale is
constrained by the asymptotic condition
$L^{u-d} (Q^2 \rightarrow \infty) = - \,\Delta \Sigma^{u-d}$,
which is a necessary consequence of QCD
evolution \cite{WN08},\cite{Thomas08}.)
However, the discrepancy between his results and the recent lattice
QCD predictions seems more than qualitative.

In any case, our semi-phenomenological analysis, which is consistent
with the empirical information and/or the lattice QCD data for
$J^u$ and $J^d$, indicates that $L^u - L^d$ remains fairly large
and negative even at the low energy scale of nonperturbative
QCD. If this is confirmed, it is a serious challenge to any low energy
models of nucleon, since they must now explain small
$\Delta \Sigma^Q$ and large and negative $L^{u-d}$ {\it simultaneously}.
The refined cloudy bag model of Thomas and Myhrer obviously fails
to do this job, since it predicts $2 \,L^u \simeq 0.64$ and
$2 \,L^d \simeq - \,0.03$
at the model scale. (See Table.1 of \cite{Thomas09}. Shown in this
table should be $2 \,L^u$ and $2 \,L^d$ not $L^u$ and $L^d$.)
Is there any low energy model which can pass this examination ?
Interestingly, the CQSM can explain
both of these peculiar features of the nucleon observables.
It has been long known that it can explain very small
$\Delta \Sigma^Q$ ($\Delta \Sigma^Q \simeq 0.35$ at the model scale)
due to the very nature of the model \cite{WY91},\cite{BEK88}.
Besides, its prediction for
$L^{u-d}$ given in \cite{WT05}, i.e. $L^{u-d} \simeq - \,0.327$
at the model scale,
perfectly matches the conclusion obtained in the present
semi-empirical analysis.

\vspace{10mm}
\noindent
\begin{large}
{\bf 6. Concluding remarks}
\end{large}

\vspace{3mm}

To conclude, the CQSM is a unique model of baryons, which has an intimate
connection with more popular Skyrme model. Although the former is an
effective quark theory, while the latter is an effective meson theory,
they share a lot of common features.
In spite of many strong similarities, a crucial difference between the
two theories was noticed already in the study of ordinary low energy
observables of the nucleon. It is a novel $1 / N_c$ correction, or
more concretely, the 1st order rotational correction, which was found to
exist within the framework of the CQSM, while it is totally missing
in the Skyrme model.
An immediate consequence of this finding is breakdown of
the so-called "Cheshire Car principle'' or the fermion-boson
correspondence.
We can show that the origin of this breakdown of fermion-boson
equivalence can eventually be traced back to the
{\it noncommutativity} of the two procedures, i.e. 
the {\it bosonization} and the
{\it collective quantization} of the rotational motion.
Alternatively, we can simply say that an important information
buried in the original fermion theory is lost in the process of
approximate bosonization. (See \cite{Waka96} for more detail.)
After all, the fact is that one is an effective quark (fermion)
theory, while the other is an effective pion (meson) theory
in $3+1$ dimension.

Superiority or wider applicability of the CQSM over the Skyrme
model becomes even more transparent if one extends the object
of research from low energy observables to the internal partonic
structure of the nucleon (or more generally of any baryons).
Since the parton distribution functions measure non-local
light-cone correlation between quarks (and gluons) inside the
nucleon, there is no way to describe them within the framework
of effective meson theories like the Skyrme model.
In contrast, this is just the place where
the potential power of the CQSM manifest most dramatically.
In this talk, we have shown, through several
concrete examples, that the CQSM provide us with an excellent
tool for theoretically understanding the nonperturbative aspect
of the internal partonic structure of the nucleon.
In particular, we have given a very plausible solution to the
longstanding ``nucleon spin problem''. We strongly believe that
the proposed solution to this famous puzzle is already
close to the truth, and it will be confirmed by experiments
to be carried out in the near future.

\vspace{5mm}
\noindent
\begin{large}
{\bf Acknowledgments}
\end{large}

\vspace{3mm}
I would like to express my sincere thanks to the hospitality of
the organizers, Profs. M.~Rosina, B.~Golli and S.~Sirca during
the workshop. Warm atmosphere and lively discussion at the
workshop are greatly acknowledged. More detailed description
of the material presented here can be found in \cite{Waka09}.


\end{document}